\documentclass{aa}  

\usepackage[switch]{lineno}
\usepackage{graphicx}
\usepackage{txfonts}
\usepackage{graphicx}
\usepackage{subfigure}
\usepackage{amsmath}
\usepackage{epsfig}

\newcommand{\imgdir}{./images}
\newcommand{\refsec}[1]{Section~\ref{#1}}
\newcommand{\reffig}[1]{Figure~\ref{#1}}
\newcommand{\refeq}[1]{Equation~(\ref{#1})}

\newcommand{\BD}{tiny-halo}
\newcommand{\Hdom}{halo-dominated}

\newcommand{\TNG}{IllustrisTNG}

\newcommand{\jMRe}{$j_\star$-$M_\star$-$R_e$}
\newcommand{\jM}{$j_\star$-$M_\star$}

\begin{document} 
%\linenumbers 
   \title{The Physical Origin of the Mass-Size Relation and Its Scatter of Disk Galaxies}

%   \subtitle{I. Overviewing the $\kappa$-mechanism}
    \titlerunning{The Mass-Size-Angular Momentum Relation}
    \authorrunning{Du+}

   \author{Min Du \inst{1} \fnmsep\thanks{E-mail: dumin@xmu.edu.cn},
          Hong-Chuan Ma\inst{1},
          Wen-Yu Zhong\inst{1},
          Luis C. Ho\inst{2, 3},
          Shihong Liao\inst{4}, 
          \and
          Yingjie Peng\inst{2, 3}
          }

   \institute{
            $^1$ Department of Astronomy, Xiamen University, Xiamen, Fujian 361005, China \\
            $^2$ Kavli Institute for Astronomy and Astrophysics, Peking University, Beijing 100871, China \\
            $^3$ Department of Astronomy, School of Physics, Peking University, Beijing 100871, China\\
            $^4$ Key Laboratory for Computational Astrophysics, National Astronomical Observatories, Chinese Academy of Sciences, Beĳing 100101, China
             }

%   \date{Received September 15, 1996; accepted March 16, 1997}

% \abstract{}{}{}{}{} 
% 5 {} token are mandatory
 
  \abstract
   {}
   {This study investigates the intricate interplay between internal (natural) and external (nurture) processes in shaping the scaling relationships of specific angular momentum (\(j_\star\)), stellar mass (\(M_\star\)), and size of disk galaxies within the \TNG\ simulation. }
   {Utilizing a kinematic decomposition of simulated galaxies, we focus on galaxies with tiny kinematically inferred stellar halos, indicative of weak external influences. The correlation among mass, size, and angular momentum of galaxies is examined by comparing simulations with observations and the theoretical predictions of the exponential hypothesis.}
   {Galaxies with tiny stellar halos exhibit a large scatter in the \jM\ relation, which suggests that it is inherently present in their initial conditions. The analysis reveals that the disks of these galaxies adhere to the exponential hypothesis, resulting in a tight fiducial \jM-scale length (size) relation that is qualitatively consistent with observations. The inherent scatter in \(j_\star\) provides a robust explanation for the mass-size relation and its substantial variability. Notably, galaxies that are moderately influenced by external processes closely adhere to a scaling relation akin to that of galaxies with tiny stellar halos. This result underscores the dominant role of internal processes in shaping the overall \jM\ and mass-size relation, with external effects playing a relatively minor role in disk galaxies. Furthermore, the correlation between galaxy size and the virial radius of the dark matter halo exists but fails to provide strong evidence of the connection between galaxies and their parent dark matter halos.}
   {}

   \keywords{Galaxy evolution -- Galaxy structures -- Galaxy stellar disks -- Galaxy stellar halos -- Astronomical simulations}

   \maketitle
%
%-------------------------------------------------------------------

\section{Introduction}

The size and morphology of galaxies provide valuable insights into the formation of galaxies and the accumulation of 
stellar mass. In the standard picture of disk galaxy formation \citep[e.g.,][]{White&Rees1978, Fall&Efstathiou1980}, disk galaxies 
are believed to form as baryons cool inside dark matter haloes, which grow through gravitational instability and acquire angular 
momentum from cosmological tidal torques \citep[e.g.][]{Hoyle1951, Peebles1969, Doroshkevich1970, White1984}. According to this paradigm, the baryons inherit the same distribution of specific angular 
momentum as the dark matter, and this conservation is maintained during the cooling process, except when large spheroids form.  
In this picture, baryonic matter settles into an exponential disk in centrifugal equilibrium. The size of this disk is 
largely determined by stellar mass $M_\star$ and specific angular momentum $j_\star$. Analytical models based on these assumptions 
have been successful in producing disk sizes for a given $M_\star$ that align reasonably well with observations
\citep[e.g.,][]{Dalcanton1997, Mo1998, Dutton2007, Somerville2008}. The size $r_e$ of a disk galaxy 
is proportional to the virial radius $r_{\rm vir}$ of its parent dark matter halo with a form as $r_e\propto \lambda r_{\rm vir}$, 
as presented by the standard framework of \citet{Mo1998}. Here the spin parameter $\lambda=j_h/(\sqrt{2}v_{\rm vir}r_{\rm vir})$ 
is a dimensionless parameter that is often used to characterize the specific angular momentum of dark matter halos $j_h$. 
$v_{\rm vir}$ is the virial velocity of the halo. 

The debate surrounding the mass and structural assembly of galaxies is still highly contested. Modern advanced cosmological hydrodynamic 
simulations, e.g., EAGLE simulations \citep{Schaye2015}, \TNG\ (TNG hereafter), SIMBA \citep{Dave2019}, and NewHorizon \citep{Dubois2021},
have successfully produced galaxies with realistic morphologies across a wide mass range, thus a powerful tool for gaining profound 
insights into physical correlations. It is well-established that galaxy sizes are related to both stellar and Active Galactic Nucleus (AGN) feedback 
\citep[summarized by][]{Somerville&Dave2015, Naab&Ostriker2017}. Therefore, some previous studies found either no correlation or only a weak 
correlation between the residual in $r_e/r_{\rm vir}$ and the halo spin parameter \citep{Teklu2015, Zavala2016, Zjupa&Springel2017, Desmond2017}. However, there 
are also clear indications that the properties of galaxies are heavily influenced by their parent dark matter halos in numerous aspects. 
For instance, \citet{Zavala2016} and \citet{Lagos2017} identified a noteworthy link between the specific angular momentum evolution of 
the dark matter and baryonic components of galaxies in EAGLE simulations. \citet{Yang2023} demonstrated that disk-dominated galaxies selected 
via kinematics in TNG and AURIGA \citep{Grand2017} reproduce a correlation between galaxy sizes and the spin parameters of their dark 
matter haloes. Similarly, \citet{Desmond2017} found a weak correlation between galaxy size and the host halo spin parameter in the 
EAGLE simulation. \citet{Liao2019} uncovered a strong correlation between sizes and host halo spin parameters for field 
dwarf galaxies in the AURIGA simulation. \citet{Zanisi2020} contended that the scatter in the galaxy-halo size relation for late-type 
galaxies could be explained by the scatter in stellar angular momentum, rather than the halo spin parameter. \citet{Jiang2019} also 
identified a weak correlation between size and spin in the VELA and NIHAO zoom-in simulations, but instead found a significant 
correlation between size and NFW halo concentration.

Especially, \citet{Du2022} showed that the TNG simulations 
\citep{Marinacci2018, Nelson2018a, Nelson2019a, Naiman2018, Pillepich2018a, Pillepich2019, Springel2018} have achieved significant success 
in replicating a $j_\star\propto M_\star^{0.55}$ relationship consistent with observations \citep[see also e.g.,][]{Rodriguez-Gomez2022}. 
Upon closer investigation of the disk galaxies from TNG, it was revealed that this scaling relation arises as a consequence of three 
physically meaningful scaling relations involving $j_\star$, $M_\star$, total mass $M_{\rm tot}$, and total specific angular momentum
$j_{\rm tot}$: (a) the $j_{\rm tot}\propto M_{\rm tot}^{0.81}$ relation deviates notably from the tidal torque theory's prediction 
of $j_{\rm tot}\propto M_{\rm tot}^{2/3}$; (b) the stellar-to-halo mass ratio consistently increases in log-log space according to 
$M_{\rm tot}\propto M_\star^{0.67}$; (c) angular momentum is approximately conserved (with a certain factor) during galaxy formation, 
i.e., $j_{\rm tot}\propto j_\star$. \citet{Du2022} suggest that the assembly of disk galaxies in the TNG simulation follows 
a consistent framework akin to that proposed by \citet{Mo1998}, but some adjustments, potentially attributable to baryonic processes, 
should be considered for a more precise understanding. 

It is crucial to disentangle the influence of external factors to comprehend the formation of galaxies in terms 
of their properties and structure. Indeed, both galaxy size and mass growth are significantly influenced by external processes, 
especially major mergers. Specifically, gas-poor mergers tend to increase galaxy 
size, whereas gas-rich mergers lead to a decrease in size \citep{Covington2008, Naab2009, Hopkins2009, Covington2011, Oser2012}. 
Consequently, the variation in galaxy size for these entities would be expected to correlate with both the frequency of mergers 
experienced by a galaxy and the gas content of those mergers, i.e., nurture. \citet{Covington2008, Covington2011} and 
\citet{Porter2014} have presented findings on the size evolution of bulge-dominated galaxies by incorporating the size growth 
observed in binary merger simulations into a semi-analytic model, exhibiting overall good agreement with observational data 
\citep[see also][]{Shankar2013}. \citet{Romanowsky&Fall2012} argued that the Hubble sequence of galaxy morphologies 
is a sequence of increasing angular momentum at any fixed mass. \citet{Obreschkow&Glazebrook2014} introduce the \jM-morphology 
relation wherein morphology is quantified by the mass fraction of bulges. \citet{Rodriguez-Gomez2022} identified a \jM-morphology relation 
in TNG, albeit with a notable degree of scatter. This picture suggests galaxy mergers, particularly ``dry" major 
mergers, can give rise to the parallel trajectory observed in the \jM\ diagram by effectively disrupting the disk 
structures of galaxies. As a consequence, earlier-type galaxies generally exhibit weaker rotational characteristics and have 
massive bulges in morphology.

Moreover, extensive research indicates that the present-day profile of a galactic disk is not primarily determined 
by the initial conditions, even in the absence of mergers. Simulations pertaining to the formation of disk galaxies 
consistently reveal that the distribution of stellar birth radii often exhibits substantial deviations from an exponential
profile \citep{Debattista2006, Roskar2008a, Roskar2012, Minchev2012, Berrier&Sellwood2015, Herpich2015}. 
This deviation occurs because stars do not remain confined to their original orbits, but exert minimal impact on the 
overall angular momentum of the galaxy disk. Both analytical arguments and 
numerical experiments have demonstrated that the angular momenta of individual disk particles are influenced by transient 
non-axisymmetric perturbations, such as spiral arms and bars, leading to a process commonly referred to as ``churning" or ``shuffling" 
\citep[e.g.,][and references therein]{Sellwood2014}. This phenomenon is commonly known as radial migration which can significantly 
change the profile of a galactic disk.

Isolating the internal and external processes can be a key to uncovering the underlying mechanisms of the assembly of 
galaxies. This duality is underscored in studies of \citet{Du2020} and \citet{Du2021}, where the authors employ a fully automatic 
kinematical method to decompose the kinematic intrinsic structures of TNG galaxies \citep{Du2019}. The mass ratio of kinematically 
derived stellar halos is sensitive to external impacts and thus can be used to quantify the effect of external processes in 
galaxies (\refsec{sec:sample}). The conceptual significance of the ``nature-nurture'' framework, within the context of 
internal versus external factors, is illustrated in \refsec{sec:samplephy}. Galaxies that have experienced minimal external 
effects give the fiducial \jM-size relation, which is detailed in Sections \ref{sec:main} and \ref{sec:jMsize}. This scaling 
relation is primarily governed by universal and natural physical processes, while nurture plays a minor role. 
In \refsec{sec:summary}, we summarize the result.

\section{Sample selection and data extraction}\label{sec:sample}

\subsection{The IllustrisTNG Simulation}

The TNG Project %\citep{Marinacci2018, Naiman2018, Nelson2018a, Nelson2019a, Nelson2019b, Pillepich2018a, Pillepich2019, Springel2018} 
is a suite of cosmological simulations run with the moving-mesh code {\tt AREPO} \citep{Springel2010, Pakmor2011, Pakmor2016} that 
utilized gravo-magnetohydrodynamics and incorporates 
a comprehensive galaxy model \citep[][]{Weinberger2017, Pillepich2018a}. The TNG50-1 run within the TNG 
suite has the highest resolution, consisting of $2 \times 2160^3$ initial resolution elements in a 
comoving box of approximately 50 Mpc. This corresponds to a baryon mass resolution of 
$8.5 \times 10^4 M_\odot$ and a gravitational softening length for stars of about 0.3 kpc 
at redshift z = 0. Dark matter particles are resolved with a mass of $4.5 \times 10^5 M_\odot$, 
while the minimum gas softening length reaches 74 comoving parsecs. These resolutions enable the 
accurate reproduction of the overall kinematic properties of galaxies with stellar masses greater than 
or equal to $10^9 M_\odot$ \citep{Pillepich2019}. 

The identification and characterization of galaxies within the simulations are performed using the 
friends-of-friends \citep{Davis1985} and {\tt SUBFIND} algorithms  \citep{Springel2001}. Resolution 
elements including gas, stars, dark matter, and black holes that belong to an individual galaxy 
are gravitationally bound to its host subhalo. All galaxies in our sample are rotated to the face-on 
view based on the stellar angular momentum to measure properties accurately.

\begin{figure}[ht!]
\begin{center}
\includegraphics[width=0.45\textwidth]{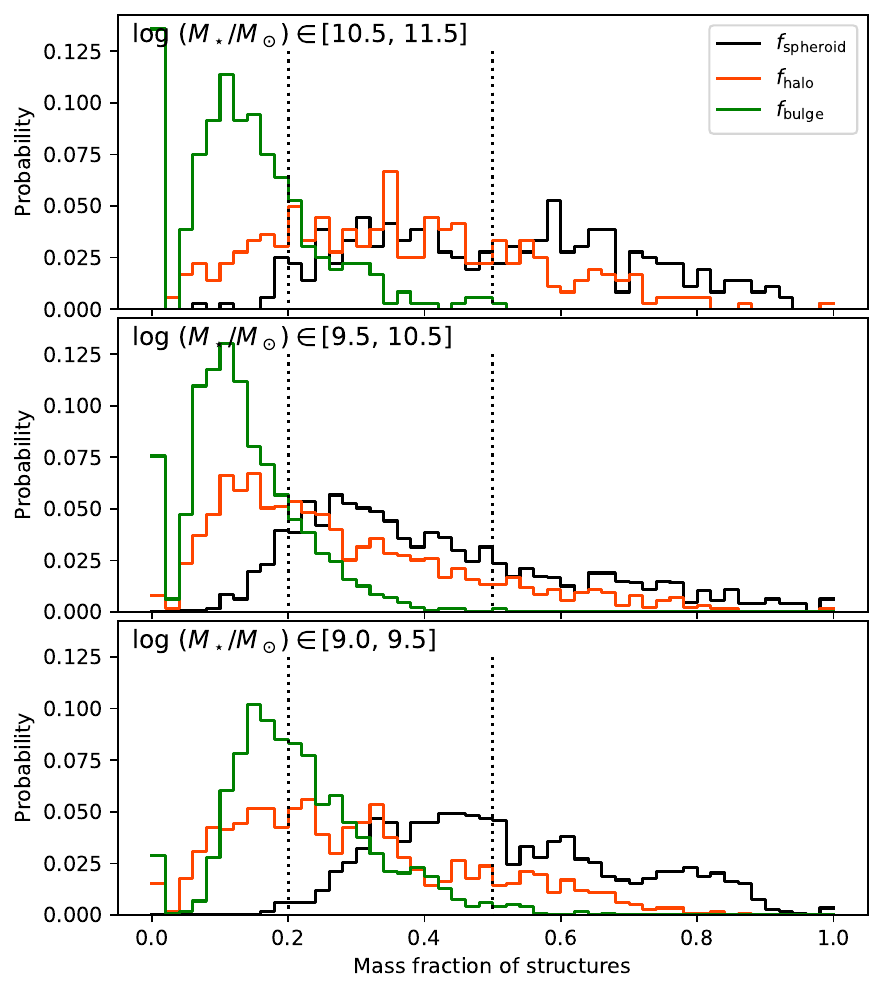}
\caption{The distribution of the mass fraction of kinematic structures in TNG50 galaxies. The bar heights are normalized their sum to 1. 
Then we classify galaxies into tiny-halo galaxies, halo-subdominated galaxies, and halo-dominated galaxies 
by $f_{\rm halo}\leq0.2$,  $0.2 < f_{\rm halo}\leq 0.5$, and $f_{\rm halo}> 0.5$, respectively. The criterion 
$f_{\rm halo}=0.2$ and $0.5$ are marked by the black vertical lines. The mass fraction of spheroids $f_{\rm spheroid}$ 
is equal to $f_{\rm bulge} + f_{\rm halo}$. From top to bottom, we show the 
galaxies with stellar mass ${\rm log}\ (M_\star/M_\odot) \in [10^9, 10^{11.5}]$. The galaxies with ${\rm log}\ (M_\star/M_\odot) \in [9.5, 10.5]$ are 
the most representative sample.
\label{fig:fstr}}
\end{center}
\end{figure}

To determine the positions of galaxies, we employ a centering method that places them at the location corresponding to the 
minimum gravitational potential energy in all measurements. The bulk velocity of all particles is subtracted. All quantities 
presented in this paper are computed using 
the full complement of particles associated with galaxies and subhalos, encompassing all gravitationally bound particles 
identified through the {\tt SUBFIND} algorithm. We refrain from imposing any constraints on the radial extent of galaxies 
when deriving their comprehensive properties. The units of length, $j$, and mass use kpc, kpc$\cdot$km s$^{-1}$, and $M_\odot$, 
respectively, over the paper. `log' always represents the logarithm with base 10.

\subsection{Physical meaning of kinematic structures: data extraction and galaxy classification}\label{sec:samplephy}

We recently developed an automated method called {\tt auto-GMM} to efficiently decompose the structures of 
simulated galaxies based on their kinematic phase space properties \citep{Du2019, Du2020}. This method used 
the {\tt GaussianMixture} Module (GMM) of the {\tt Python scikit-learn} package to cluster the three-dimensional phase 
space of dimensional parameters that quantify circularity, 
binding energy, and non-azimuthal angular momentum \citep[][]{Domenech-Moral2012} into distinct structures. Such 
kind of kinematic method has become a standard way to decompose galaxies accurately 
\citep[see similar attempts in][]{Obreja2018a, Zana2022, Proctor2024}. We successfully identified cold disk, warm disk, bulge, and 
stellar halo structures of TNG galaxies using {\tt auto-GMM} \citep{Du2020}. The overall disk and spheroidal structures are obtained by summing the 
stars from the cold+warm disks and the bulge+stellar halo, respectively. 
%These classification criteria were inferred heuristically from statistical analyses of disk galaxies in the TNG100 simulation. 
Notably, stars within kinematically derived disks are predominantly characterized by strong rotation, exhibiting 
a mass-weighted average circularity $\langle j_z/j_{\rm c}\rangle>0.5$, where $j_z$ and $j_{\rm c}$ are 
the azimuthal and circular angular momentum, respectively. Conversely, the kinematically derived stellar halos share a 
similar weak rotation ($\langle j_z/j_{\rm c}\rangle<0.5$) with bulges but possess looser binding stars than bulges. 
It's important to emphasize that this decomposition method does not assume that the disk of a galaxy follows 
an exponential profile, nor does it presuppose that the bulge follows a S$\acute{e}$rsic profile. 

In \citet{Du2021}, we propose the ``nature-nurture'' picture to understand the structures of galaxies and their evolution.
In this paper, ``nature'' is equivalent to internal processes, while ``nurture'' is equivalent to external processes.
By successfully identifying kinematic structures within galaxies, it becomes possible to establish 
connections between these structures and either nature (internal) or nurture (external) physical processes.
The early phase at redshifts $z>2$, characterized by chaotic physical processes and gas accretion in the host dark matter halo and protogalaxy, 
is regarded as the one aspect of galaxy nature. In the later phase, there is no doubt that long-term evolution belongs to 
nature in the absence of significant mergers. Consequently, the nature of galaxies substantially contributes to the formation 
of both kinematic bulges and disk structures, as demonstrated by \citet{Du2021}. In contrast, only kinematic stellar 
halos are strongly linked to external events, primarily mergers but not exclusively, representing the 
``nurture" aspect. 

In this study, our primary focus is to elucidate the influence of internal and external processes on the scaling relations 
of \jM\ and the size of galaxies from TNG50. We utilize galaxies within the stellar mass range of $10^{9}-10^{11.5} M_\odot$ 
from the TNG50-1 simulation. \reffig{fig:fstr} presents the distribution of the mass fractions of kinematic structures 
in three stellar mass ranges from top to bottom, as detailed in 
\citet{Du2021}\footnote{The data of kinematic structures in TNG galaxies are publicly accessible at https://www.tng-project.org/data/docs/specifications/\#sec5m}.
We proceeded to categorize galaxies into three groups based on their stellar halo mass fractions $f_{\rm halo}$, as follows:
\begin{itemize}
    \item {\it Tiny-halo galaxies}: $f_{\rm halo} \leq 0.2$ select 997 galaxies. These galaxies are robustly classified as disk galaxies in terms of morphology and can be considered to have formed via internal processes, largely unaffected by mergers and other environmental factors. They serve as the physical basis for the fiducial scaling relations.
    \item {\it Halo-subdominated galaxies: $0.2 < f_{\rm halo} \leq 0.5$} select 1369 galaxies. The morphological classification of such galaxies is challenging. The existence of a massive stellar halo is a sign that such a galaxy has experienced somewhat external effects. It thus may lead to a scatter in any fiducial scaling relation originating from internal processes.
    \item {\it Halo-dominated galaxies}: $f_{\rm halo} > 0.5$ select 442 galaxies, indicating elliptical galaxy morphology. For these galaxies, any fiducial scaling relation resulting from internal processes may have been disrupted or substantially altered due to the pronounced influence of external processes.
\end{itemize}
In comparison with the ex-situ mass fraction measured in \citet{Rodriguez-Gomez2015}, the use of $f_{\rm halo}$ is more 
convenient, as it eliminates the need to account for variations in the strength, orbits, and frequency of mergers and 
close tidal interactions. The mass fraction of spheroids $f_{\rm spheroid}$ in 
each galaxy can be computed simply as $f_{\rm bulge} + f_{\rm halo}$.

\begin{figure*}[ht!]
\begin{center}
\includegraphics[width=\textwidth]{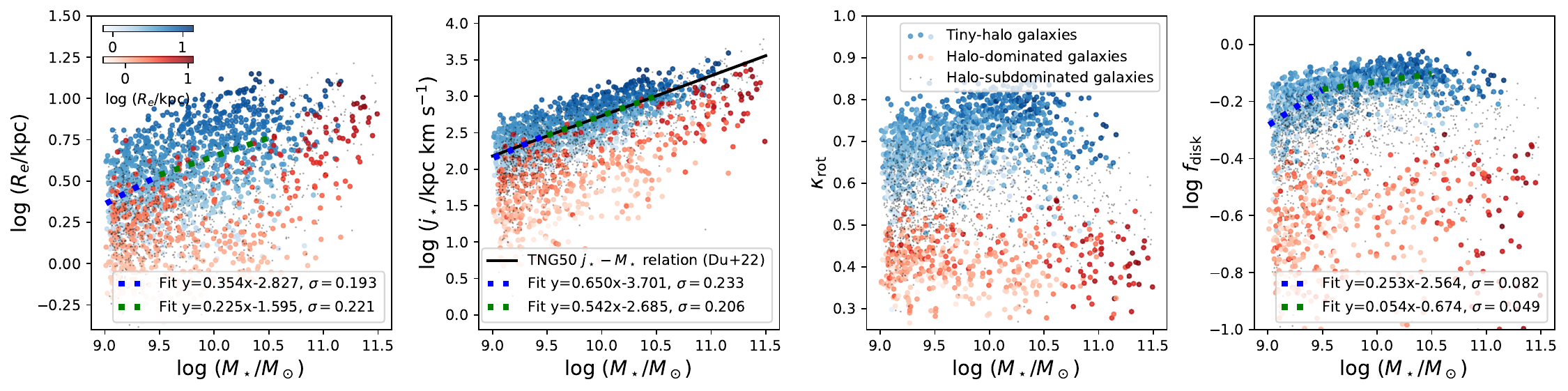}
\caption{The scaling relations of \BD\ (blue dots), halo-subdominated (gray dots), and \Hdom\ (red dots) galaxies.  
From left to right, we show $M_\star$-$R_e$ relation, $j_\star$-$M_\star$ relation, rotation $\kappa_{\rm rot}$, 
and the mass fraction of disk structures $f_{\rm disk}$. The deepness of red and blue colors represents the 
${\rm log}\ (R_e/{\rm kpc})$ in all panels, showing that larger galaxies have relatively larger $j_\star$ 
for a given stellar mass. The green triangle symbols show the observational result 
from \citet{Pina2021}. We perform the linear fitting for galaxies in two mass ranges ${\rm log}(M_\star/M_\odot) \in [9.5, 11.5]$ 
(green dotted lines) and $[9., 9.5]$ (blue dotted lines), respectively. The \jM\ relation of disk galaxies from \citet{Du2022} 
is overlaid in black in the second panel.
\label{fig:jM}}
\end{center}
\end{figure*}

\section{The fiducial \lowercase{$j_\star$}-$M_\star$-$\lowercase{h}_R$ plane of bulge/disk-dominated galaxies from their exponential nature}\label{sec:main} 

It is well known that more massive galaxies have larger sizes, although the scatter is large at given $M_\star$ 
\citep[e.g.,][]{Shen2003, FernandezLorenzo2013, Lange2015, Munoz-Mateos2015}. TNG has successfully reproduced the relation between stellar mass and half-mass radius $R_e$ 
within observational uncertainties \citep[e.g.,][see also the left-most panel of \reffig{fig:jM}]{Genel2018, Huertas-Company2019, Rodriguez-Gomez2019}.
But the substantial scatter in galaxy sizes, ranging from 1 kpc to more than 10 kpc, and in $j_\star$ \citep{Du2022, Fall2023} continue
to pose a perplexing challenge. In this study, we isolate the influence of internal processes on the mass-size relation and $j_\star$ by selecting 
the \BD\ galaxies. The effect of nurture then is shown by comparing halo-subdominated and halo-dominated galaxies with their counterparts with 
tiny stellar halos. 

\subsection{The large scatter of the $j_\star$-$M_\star$ relation}

Conducting a comparative study is crucial to differentiate between the effects of internal and external processes on galaxy 
evolution and the large scatter of the \jM\ relation. 
In the second panel of Figure \ref{fig:jM}, we show the $j_\star$-$M_\star$ relation for the three types of galaxies defined 
in \refsec{sec:sample}, e.g., \BD\ galaxies (blue dots), halo-subdominated galaxies (gray dots), and \Hdom\ galaxies (red dots). 
Tiny-halo galaxies in the absence of mergers exhibit the almost exactly same \jM\ relation (the blue and green dotted lines fitted in different 
mass ranges) to disk galaxies selected based on the relative importance of cylindrical rotations ($\kappa_{\rm rot}\geq 0.5$) as described in 
\citet{Du2022}, corresponding to the black solid line. Specifically, these galaxies follow a scaling relation of $j_\star\propto M_\star^{0.55}$. 
This finding suggests 
that the wide scatter observed around the $j_\star\propto M_\star^{0.55}$ relation exists regardless of whether external processes 
have played a significant role in their evolution. Moreover, it is evident that halo-subdominated and \Hdom\ galaxies generally possess 
smaller $j_\star$ values for a given $M_\star$. It is not surprising that mergers induce the increase of mass as well as the decrease of 
angular momentum via destroying disky structures.

\begin{figure}[ht!]
\begin{center}
\includegraphics[width=0.48\textwidth]{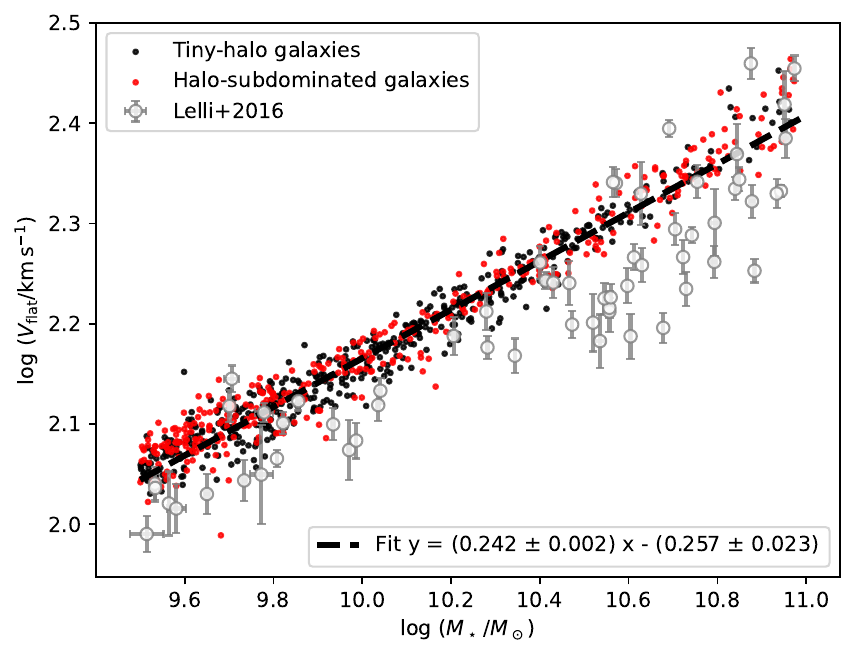}
\caption{The Tully-Fisher relation of galaxies in TNG (black and red dots) and observations. The $v_{\rm flat}$ of TNG galaxies is measured by averaging 
the flat part ($0.05-0.2 r_{\rm vir}$) of the rotation curve. The observations are adopted from \citet{Lelli2016} assuming 
the mass-to-light ratio $\Gamma=0.5 M_\odot/L_\odot$ based on the IMF suggested by \citet{Kroupa2001}.
\label{fig:TF}}
\end{center}
\end{figure}

The extensive scatter observed in the \jM\ relation of \BD\ galaxies exhibits a distinct correlation with the galaxy size, as indicated 
by the depth of the blue color in the plot. Furthermore, our analysis does not reveal a significant correlation between rotation and 
$R_e$ in these galaxies, as demonstrated in the third and fourth panels of Figure \ref{fig:jM}. The substantial scatter in the \jM\ 
relation primarily stems from the significant variation in galaxy size, driven by internal processes, as elucidated in Sections 
\ref{sec:jMhR} and \ref{sec:jMsize}. It is not surprising that both halo-subdominated and halo-dominated galaxies exhibit a clear deviation 
towards lower $j_\star$ compared to the \jM\ relation of \BD\ galaxies, which is consistent with the prediction of the so-called 
\jM-morphology relation \citep{Sweet2018, Obreschkow&Glazebrook2014, Rodriguez-Gomez2022}. 
This divergence is linked to their noticeably weaker rotation shown in the third and fourth panels.
But the scatter of $j_\star$ from nature plays a more important role. We thus can conclude that a pronounced 
scatter in the \jM\ relation is inherently present in the initial conditions of galaxies, shaping the observed \jM\ relation and 
mass-size relation in the local Universe. On the other hand, external influences induce a systematic offset, further 
contributing to this inherent variation.

\begin{figure*}[ht!]
\begin{center}
\includegraphics[width=0.48\textwidth]{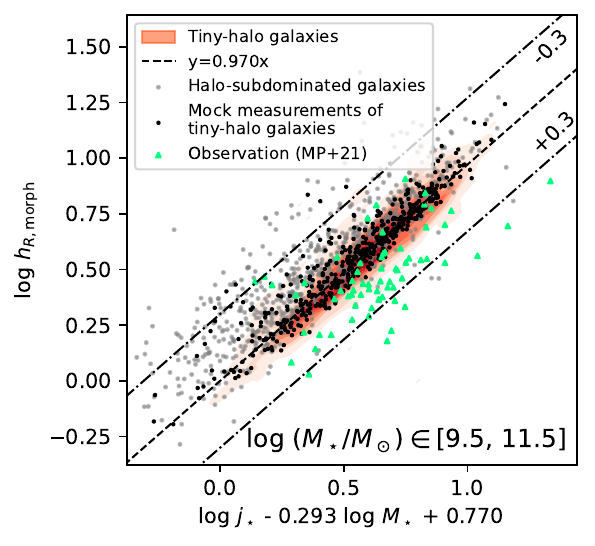}
\includegraphics[width=0.48\textwidth]{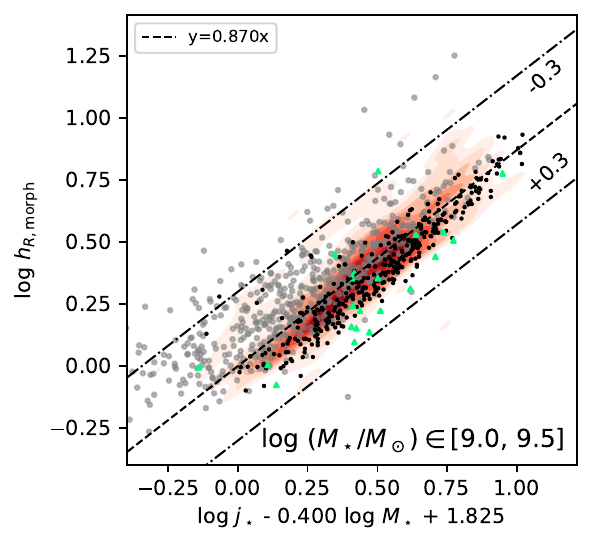}
\caption{The \jM-$h_{\rm R, morph}$ relation comparing simulations with observations. The units of $h_{R, {\rm morph}}$, $j_\star$, 
and $M_\star$ use kpc, kpc$\cdot$km/s, and $M_\odot$, respectively. The left and right panels show galaxies in two mass ranges 
${\rm log}\ (M_\star/M_\odot) \in [9.5, 11.5]$ and $\in [9.0, 9.5]$. The fiducial \jM-$h_{\rm R, morph}$ relation of \BD\ galaxies, 
where $h_{\rm R, morph}$ represents the scale length of the disk component defined by morphology, is visualized using kernel density 
estimation (KDE) map. For convenience, we adjust the surface fitting to align with $y=x$. We perform mock measurements of \BD\ galaxies 
based on \refeq{eq:js} (black dots). The gray points show the halo-subdominated galaxies for comparison. The observational data points of disk galaxies
from \citet{Pina2021} are represented by green triangles. 
%The black circles highlight the subset of the sample that exhibits a significant offset of +0.6 dex along the $x$-axis. 
The dot-dashed lines represent the cases exhibiting offsets of $\pm 0.3$ dex. 
\label{fig:fp3}}
\end{center}
\end{figure*}

\subsection{The fiducial \jM-$h_R$ relation in nature satisfying the exponential hypothesis}\label{sec:jMhR}

In a realistic scenario of the exponential hypothesis, galaxies consist of both spheroidal and disk components. 
The angular momentum of a galaxy is largely determined by the mass fraction of its disk components, denoted as 
$f_{\rm disk}$, as spheroidal components exhibit little or no rotation. Here a disk structure includes thin (cold) 
and thick (warm) disk components. We then have
\begin{equation}
\begin{split}\label{eq:js0}
    j_{\star, {\rm theory}} &= f_{\rm disk}j_{\rm disk} \\%+ f_{\rm sph} j_{\rm sph} \\
            &= \frac{1}{M_\star} \int 2 \pi R^2 \Sigma_\mathrm{disk}(R) v_\phi(R) \mathrm{d}R. \\
\end{split}
\end{equation}
$v_\phi$ and $\Sigma_\mathrm{disk}$ are the cylindrical rotation velocity and surface density at cylindrical radius $R$, respectively. 
Exponential disks exhibit a simple surface density profile described as $\Sigma_\mathrm{disk}=\Sigma_\mathrm{0, disk}\exp{(-R/h_{R, {\rm theory}})}$. 
The central surface density $\Sigma_\mathrm{0, disk}$ can be expressed as $f_\mathrm{disk}M_\star/(2\pi h_{R, {\rm theory}}^2)$. 
$\int R^2 \exp{(-R/h_{R, {\rm theory}})} \mathrm{d}R$ integrates from 0 to infinity resulting in $2h_{R, {\rm theory}}^3$.
The accuracy of estimating $j_{\star, {\rm theory}}$ using \refeq{eq:js0} hinges on the precise characterization of disk structures by 
$h_{R, {\rm theory}}$, $f_{\rm disk}$, and their rotation curves. We define the factor $\epsilon=\langle v_\phi/v_{\rm flat}\rangle$ 
to quantify the deviation of $v_\phi$ from the flat rotation curve with a velocity $v_{\rm flat}$.
$\epsilon$ thus has a similar physical meaning to the circularity $\langle j_z/j_{\rm c}\rangle$. 
%This is derived from the relation $M_{\rm disk} = f_\mathrm{disk}M_\star = \Sigma_\mathrm{0, disk} \int 2 \pi R e^{-R/h_R} \mathrm{d}R = 2 \pi h_R^2\Sigma_\mathrm{0, disk}$, where $\Sigma_\mathrm{0, disk}$ represents the central surface density.
Through a simple derivation, \refeq{eq:js0} gives 
\begin{equation}
\begin{split}\label{eq:js}
    j_{\star, {\rm theory}} &=  \frac{2 \pi \epsilon \Sigma_\mathrm{0, disk} v_{\rm flat}}{M_\star} \int R^2 e^{-R/h_{R, {\rm theory}}} \mathrm{d}R \\
%            &= v_{\rm flat} \frac{f_{\rm disk}}{h^2} 2 h_R^3 \\
            &= 2\epsilon f_\mathrm{disk} v_{\rm flat} h_{R, {\rm theory}} 
%            &\simeq f_\mathrm{d} h \sqrt{M_\star/R_e}
\end{split}
\end{equation}
%$j_{\rm sph}$ is the specific angular momentum of spheroidal components in galaxies that are close to 0. 
This equation is physically robust in cases where galaxies satisfy the exponential hypothesis. It thus has been commonly 
used as an approximation of $j_\star$ \citep[e.g.,][]{Fall1983, Mo1998}, via measuring $f_{\rm disk}$ in morphology and 
making corrections by adding asymmetric drift. 

It should be noted that \refeq{eq:js} holds when we can accurately measure the mass fraction of disks that 
possess strong rotation and conform to an exponential distribution. Moreover, the TF relation 
$M_\star\propto v_{\rm flat}^\alpha$ is generally tightly satisfied where the gaseous component is  
negligible in the local Universe. Observations give a consistent result in previous studies that 
$\alpha$ varies between 3 and 4 \citep[e.g.,][]{Noordermeer&Verheijen2007, Avila-Reese2008, Gurovich2010, Zaritsky2014, Bradford2016, Papastergis2016, Lelli2019}. 
%Careful measurement of the baryonic Tully-Fisher relation in IllustrisTNG simulations presented in \citet{Goddy2023} gives 
%${\rm log} M_b = 3.47\ {\rm log} v_{\rm flat} + 2.61$ in the case that all observational effects are ignored 
%\citep[see also][]{Faucher2023}. 
The TF relation we measured in TNG50 \BD\ galaxies (black dots in \reffig{fig:TF}) gives 
${\rm log}\ (v_{\rm flat}/{\rm km\ s^{-1}}) = (0.242\pm 0.002)\ {\rm log}\ (M_\star/M_\odot) - (0.257\pm 0.023)$ 
that is consistent with the observation from \citet{Lelli2016} (gray dots with error bars).%, see \citet{Faucher2023} and \citet{Goddy2023} also for the measurement of the baryonic Tully-Fisher relation. 
We then have the fiducial \jM-size relation based on the theory of the exponential hypothesis
%The Tully-Fisher relation was measured using a large data set of $\sim1600$ local disk galaxies compiled by Courteau et al. (2006) using I- and K-band velocity-luminosity relation i. The slope gives $\sim 0.3$. (Bradford et al. 2016; Lelli et al. 2019).
\begin{equation}\label{eq:jMhRphy0}
    \begin{split}
    \mathrm{log}\ &(h_{R, {\rm theory}}/{\rm kpc}) \simeq \\
        &\mathrm{log}\ (j_{\star, {\rm theory}}/ {\rm kpc\ km\ s^{-1}}) -  0.24\ \mathrm{log}\ (M_\star/M_\odot) + C_0. % - \mathrm{log}\ (\epsilon f_{\rm disk}).
%        &\mathrm{log}\ (j_{\star, {\rm theory}}/ {\rm kpc\ km\ s^{-1}}) -  0.28\ \mathrm{log}\ (M_\star/M_\odot) + C. % - \mathrm{log}\ (\epsilon f_{\rm disk}).
    \end{split}
\end{equation}
The constant part $C_0$ is $-\mathrm{log}\ (2\epsilon f_{\rm disk}) - C_{\rm TF}$ where $C_{\rm TF}=-0.26$ is the zero point of the 
TF relation. $C_0$ is nearly constant around 0.3 estimated by $f_{\rm disk}\sim 0.7$ and $\epsilon \sim 0.7$. %\sim \mathrm{log}\ (2\epsilon f_{\rm disk}
After considering the correction from $f_{\rm disk}$, the right-most panel of \reffig{fig:jM} gives ${\rm log}\ f_{\rm disk}=0.054\ {\rm log}\ (M_\star/M_\odot)-0.674$. Then we have 
\begin{equation}\label{eq:jMhRphy}
    \begin{split}
    \mathrm{log}\ &(h_{R, {\rm theory}}/{\rm kpc}) \simeq \\
%        &\mathrm{log}\ (j_{\star, {\rm theory}}/ {\rm kpc\ km\ s^{-1}}) -  0.23\ \mathrm{log}\ (M_\star/M_\odot) + C. % - \mathrm{log}\ (\epsilon f_{\rm disk}).
         &\mathrm{log}\ (j_{\star, {\rm theory}}/ {\rm kpc\ km\ s^{-1}}) -  0.29\ \mathrm{log}\ (M_\star/M_\odot) + C_1 % - \mathrm{log}\ (\epsilon f_{\rm disk}).
    \end{split}
\end{equation}
where $C_1=-\mathrm{log}\ (2\epsilon)+0.94$.  $\epsilon$ varies in the range of $0.85-1.0$ and $0.5-0.85$ for cold and warm disks, 
respectively, defined by the kinematical 
method in \citet{Du2020}. It is roughly constant around $0.7$ estimated by the relative mass fraction of cold and warm disks. 
A dynamically hotter disk has a smaller $\epsilon$. Then $C_1$ is about 0.79. The upper and lower 
limits can be 0.94 and 0.64 in the case of $\epsilon=0.5$ and $\epsilon=1$, respectively. Such a theoretical \jM-$h_R$ relation relies 
on the assumption that disks accurately satisfy the exponential profile and bulges have zero rotation. 
It is worth emphasizing that the correction of ${\rm log} f_{\rm disk}$ is non-negligible, which may induce a deviation 
of $C_1-C_0\sim 0.5$ dex when the difference in the factor of the ${\rm log} M_\star$ is ignored.

We examine whether \BD\ galaxies in TNG50 abey the theoretical \jM-$h_R$ relation (\refeq{eq:jMhRphy}). 
We first perform a surface fitting in the 3-dimensional (3D) space using $j_\star$, $M_\star$, and $h_{R, {\rm morph}}$ for \BD\ galaxies in 
two mass ranges ${\rm log}\ (M_\star/M_\odot) \in [9, 9.5]$ and $\in [9.5, 11.5]$. Figure \ref{fig:fp3} then shows the fitting 
results in a 2D way which is convenient to compare with \refeq{eq:jMhRphy}. $h_{R, {\rm morph}}$ is extracted using a 1D two-component 
(S$\acute{e}$rsic bulge + exponential disk) morphological decomposition that has been widely used in observations. 
We here use a 1D bulge-disk decomposition to simplify the analysis, as the face-on surface density map of galaxies is exactly 
known in simulations. 

The fitting result of the fiducial \jM-$h_R$ relation of \BD\ galaxies with 
${\rm log}\ (M_\star/M_\odot) \in [9.5, 11.5]$ in TNG50 simulation gives
\begin{equation}
\begin{split}
    \mathrm{log}\ &h_{R, {\rm morph}} = (0.970\pm 0.020)\ [\mathrm{log}\ j_\star \\ 
                    & -(0.293\pm 0.015)\ \mathrm{log}\ M_\star + (0.770\pm 0.123)]
\end{split}
\end{equation}
Where the units of $h_{R, {\rm morph}}$, $j_\star$, and $M_\star$ use kpc, kpc$\cdot$km/s, and $M_\odot$, respectively. The left and 
right parts of this equation are used as the $y$ and $x$ axes, respectively, in \reffig{fig:fp3}.
The fitting result matches \refeq{eq:jMhRphy} perfectly, suggesting that galaxies evolve in a natural way obeying 
the exponential hypothesis. It is worth emphasizing that the kinematic disk structures indeed show some noticeable deviations from simple exponential profiles \citep[see Figure 12 in][]{Du2022}, while the exponential hypothesis is still valid to explain the overall properties.
This result is not sensitive to stellar mass for massive galaxies but has a clear deviation in 
less-massive galaxies. The right panel of \reffig{fig:fp3} gives the fiducial \jM-$h_R$ relation 
of dwarf galaxies with ${\rm log}\ (M_\star/M_\odot) \in [9, 9.5]$
\begin{equation}
\begin{split}
    \mathrm{log}\ &h_{R, {\rm morph}} = (0.870\pm 0.024)\ [\mathrm{log}\ j_\star \\ 
                    & -(0.400\pm 0.043)\ \mathrm{log}\ M_\star + (1.825\pm 0.387)].
\end{split}
\end{equation}
The deviation with respect to the more massive galaxies is largely due to that ${\rm log}\ f_{\rm disk}$ (blue 
dashed line) has a steeper slope and a smaller intercept, as shown in the right-most panel of \reffig{fig:jM}. 

Moreover, the halo-subdominated galaxies (gray dots) exhibit a similar \jM-$h_R$ relation, albeit with a noticeable degree of 
scatter. These galaxies clearly deviate towards the left side in comparison to the fiducial \jM-$h_R$ relation depicted 
in \reffig{fig:fp3}. This deviation is likely attributed to the reduced $j_\star$ from external influences. 

In summary, disk structures of TNG galaxies do conform to the exponential hypothesis. The significant scatter in the \jM\ relation, 
as seen in \reffig{fig:jM}, which is inherent in protogalaxies or their host dark matter halos, effectively explains the 
underlying physical basis for the fiducial \jM-$h_R$ relation. Consequently, this scatter results in a wide range of 
galaxy sizes, indicating that the evolution of these galaxies has only been minimally impacted by external 
influences. External factors play a relatively minor role in shaping the overall mass-size and \jM\ scaling relationships 
in disk galaxies. Moreover, this result suggests that the effect of stellar migrations
\citep[e.g.][]{Debattista2006, Roskar2012, Berrier&Sellwood2015, Herpich2015} also has a minor effect on the 
overall properties of disk galaxies.

\subsection{The deviation of the \jM-$h_R$ relation between the TNG50 simulation and observations}\label{sec:jMhRobs}

The fiducial \jM-$h_R$ relation on the basis of the exponential hypothesis provides a reference point for galaxies that are primarily 
rotation-dominated. It provides valuable constraints on the physical model to explain the galaxy's size. In \reffig{fig:fp3}, we 
compare the observational data with the fiducial \jM-$h_R$ relation derived from TNG50. 
The observational data of $j_\star$ utilized in this study are sourced from \citet{Pina2021} (green triangles). 
It is worth emphasizing that all observed galaxies here are in close proximity, allowing for relatively reliable measurements 
of $M_\star$ and $h_{R, {\rm morph}}$ enabling a meaningful comparison. 
The estimation of the mass fraction and $h_{R, {\rm morph}}$ of disks is based on 2D bulge-disk 
decomposition conducted by \citet{Fisher&Drory2008}. This decomposition combines high-resolution Hubble Space Telescope 
imaging with wide-field ground-based imaging, which helps to minimize uncertainties of $h_{R, {\rm morph}}$ and $f_{\rm disk}$. 

Galaxies in TNG50 follow a similar trend to those in observations. But there is indeed a notable discrepancy between the 
observed galaxies of \citet{Pina2021} and the galaxies in TNG50, as illustrated in the left panel of \reffig{fig:fp3}. This discrepancy 
is smaller in less massive galaxies, as seen in the right panel.
It is evident that many galaxies exhibit an offset of approximately $>0.2$ dex to the right 
of the fiducial \jM-$h_R$ relation. %The offset even reaches up to about 0.6 dex when compared to the IFU data from \citet{Sweet2018}. 
We first investigate the uncertainty of observations to figure out the potential source of the observed offset. The determination of 
$j_\star$ in \citet{Pina2021} involves measuring the rotation curve from gas after applying a stellar-asymmetric drift correction.
Part of the measurements of $j_\star$ \citep[e.g.,][]{Romanowsky&Fall2012} employs slit spectroscopy of both starlight and ionized gas. 
According to \citet{Sweet2018}, the typical uncertainty of $j_\star$ is given by $|\Delta j_\star|/j_\star=0.05-0.1$, reaching a maximum 
of 0.32 ($\sim 0.15$ dex) for the data from \citet{Romanowsky&Fall2012}. Moreover, the uncertainty of $M_\star$ is about 0.2 dex estimated 
by the uncertainty of the mass-to-light ratio adopting a factor of $\sim 1.5$. To estimate the overall uncertainty of 
${\rm log}\ j_\star - 0.3\ {\rm log}\ M_\star$, we can use the formula $\sqrt{0.15^2+(0.3\times 0.2)^2}\approx0.16$. 
This uncertainty can partially explain the inconsistency between the observational results using \refeq{eq:js} 
and simulations. 

To quantify any potential uncertainty in the measurement of $j_\star$, we perform mock measurements based on \refeq{eq:js} (black dots 
in \reffig{fig:fp3}). The $v_{\rm flat}$ is measured by the average value of the flat part of the outer edge ($0.05-0.2 r_{\rm vir}$) 
of a rotating curve. We do not make any asymmetric drift correction which will only lead to a negligible offset toward the left side. 
It thus cannot explain the deviation between observations and TNG simulations. We have confirmed that halo-subdominated galaxies measured 
using \refeq{eq:js} follow a similar \jM-$h_R$ relation. It is likely because of the fact that the transformation from disks to stellar halos due 
to mergers generally induces a minor change on both $v_{\rm flat}$ and $h_{R, {\rm morph}}$ measured in morphological decomposition. 
It is clear that halo-subdominated galaxies (red dots) follow a similar TF relation to \BD\ galaxies (black dots), as shown in \reffig{fig:TF}.
Thus, $j_\star$ of halo-subdominated galaxies are likely to be significantly overestimated using this method, while it will not affect 
our results in this study.

The observation may not be able to reflect the physics due to poor statistics and large uncertainty. The 3D surface 
fitting of observational data from \citet{Pina2021} gives
\begin{equation}\label{eq:obsjs}
\begin{split}
    \mathrm{log}\ &h_{R, {\rm morph}} = (0.028\pm 0.089)\ \mathrm{log}\ j_\star \\ 
                    & +(0.345\pm 0.130)\ \mathrm{log}\ M_\star - (0.844\pm 0.661).
\end{split}
\end{equation} 
This result suggests that $h_{R, \text{morph}} \propto M_\star^{1/3}$, which is nearly independent of $j_\star$. The strong 
correlation that implies $h_{R, \text{morph}} \propto j_\star$ in the fiducial \jM-$h_R$ relation disappears when only the 
observational data is considered. However, there is indeed a substantial uncertainty $\pm 0.661$ dex in the constant part 
on the right side of \refeq{eq:obsjs}, indicating that no satisfactory 3D fitting results can be obtained. 
Moreover, it is worth mentioning that the deviation between simulations and an integral field spectroscopic (IFS) measurement \citep{Sweet2018} 
is even larger. \citet{Obreschkow&Glazebrook2014} have already noticed that $j_\star$ measured by \refeq{eq:js} have systematic 
variations in comparison to the IFS observations. Many observational issues should be examined in detail. We thus do not compare 
with the result of IFS measurement in this work. Consequently, we still cannot reach a robust conclusion due to the poor 
statistics and large uncertainty of the observational data.

\subsection{No strong evidence of incorrect galaxy properties in \TNG\ simulations}\label{sec:incorrect}

In the previous section, we demonstrated that the disparity between simulations and observations is predominantly due to measurement 
uncertainties. As of now, we cannot rule out the possibility that this inconsistency is a result of errors in the 
simulated galaxy properties generated by the \TNG\ simulations. Generating galaxies with a faster flat rotation curve, a smaller disk size, or 
a larger mass fraction of disk structures in simulations may solve its inconsistency with observations. 
There is no clear inconsistency in the TF relation, as shown in \reffig{fig:TF}. Though the observational findings 
by \citet{Lelli2016} (represented by open dots with error bars) are marginally lower compared to the disk galaxies in the TNG50 simulation, 
this small discrepancy only accounts for a negligible 0.05 dex deviation in the fiducial \jM-$h_R$ relation. In this section, we further delve 
into the specific properties of disk galaxies in TNG50 to elucidate the discrepancies in the \jM-$h_R$ relation observed.

\begin{figure}[ht!]
\begin{center}
\includegraphics[width=0.48\textwidth]{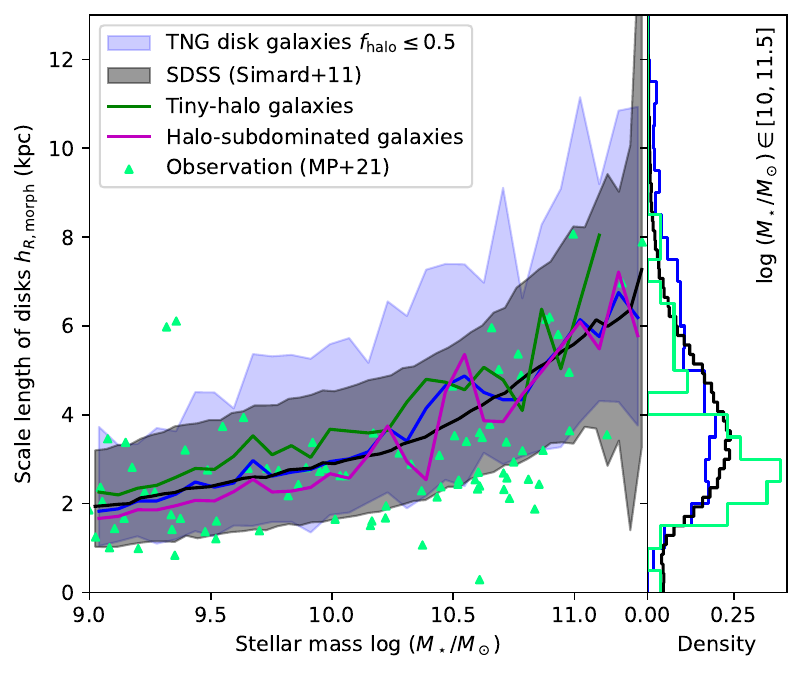}
\caption{The mass-$h_{R, {\rm morph}}$ diagram. Disk galaxies from TNG50 are represented by the blue-shaded area, while 
observed disk galaxies from SDSS DR7 are shown in the gray-shaded region. The solid profiles are the median. The shaded 
areas denote the $1\sigma$ envelope, representing the 16th and 84th percentiles. For the SDSS data, we utilized scale 
length values from \citet{Simard2011} and stellar mass data from \citet{Mendel2014}. It is apparent that SDSS galaxies 
are consistent with TNG50 disk galaxies, and the difference between halo-subdominated galaxies and \BD\ galaxies is minimal. 
The green triangles and the histograms on the right illustrate that the data utilized in \citet{Pina2021} (green) 
are biased towards smaller-sized galaxies.
% and \citet{Pina2021} (green triangles). 
\label{fig:masssize}}
\end{center}
\end{figure}

In \reffig{fig:masssize}, we show the mass-$h_R$ relation, comparing the \BD\ galaxies from TNG50 (represented by blue shaded regions) 
with those observed (triangles). Additionally, we overlay the late-type galaxies from the SDSS DR7 dataset in black. The 
selection criterion for late-type galaxies is based on a color threshold of $g-r<0.7$, as suggested by \citet{Blanton2003}. We 
adopt the scale length approximated by \citet{Simard2011} and the stellar mass provided by \citet{Mendel2014}.
It is evident that galaxies in SDSS observations (black histogram and shaded regions) demonstrate a consistent trend with galaxies 
in TNG50 simulations (blue histogram and shaded regions). The galaxies utilized in \citet{Pina2021} include a group of compact 
massive galaxies as we can see by comparing the green with the black and blue histograms on the right side of \reffig{fig:masssize}. Such 
galaxies generally have stellar masses larger than $10^{10.2} M_\odot$, but compact disks with $h_R < 2$ kpc. Such galaxies are 
uncommon. It suggests that the galaxies selected for $j_\star$ measurements may be biased towards compact galaxies, which may 
not be representative enough to draw definitive conclusions.
%For example, compact galaxies from \citet{Sweet2018} falling below the dashed lines generally significantly offset from the fiducial 
%\jM-$h_R$ relation, as marked by the black edges in \reffig{fig:fp3}. However, there are still many galaxies that are not that compact.

\begin{figure}[ht!]
\begin{center}
\includegraphics[width=0.48\textwidth]{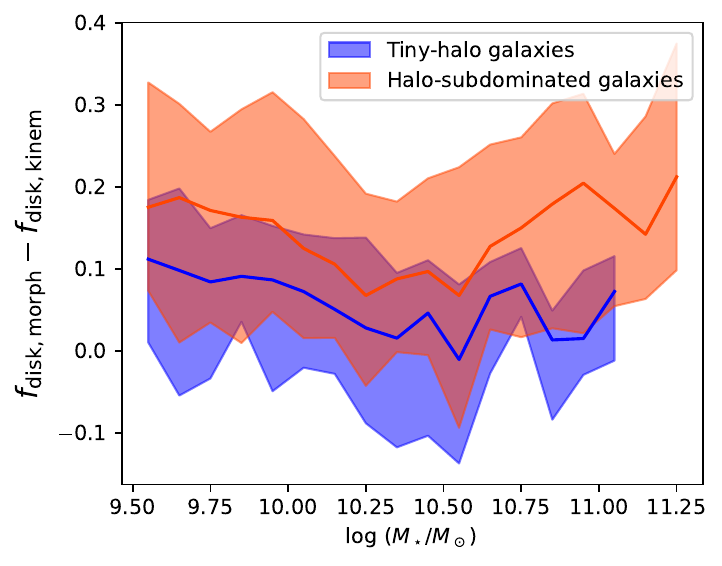}
\caption{The relative mass fraction of disk structures of \BD\ (blue) and halo-subdominated (red) galaxies, measured by morphological $f_{\rm disk, morph}$ and kinematic $f_{\rm disk, kinem}$ methods. 
The morphologically defined disks are generally 10 percent larger than those defined in the kinematical method from \citet{Du2019, Du2020}. 
The envelopes of shaded regions represent the 16th and 84th percentiles and the solid profile is the median value.
\label{fig:fdisk}}
\end{center}
\end{figure}

\begin{figure*}[ht!]
\begin{center}
\includegraphics[width=0.48\textwidth]{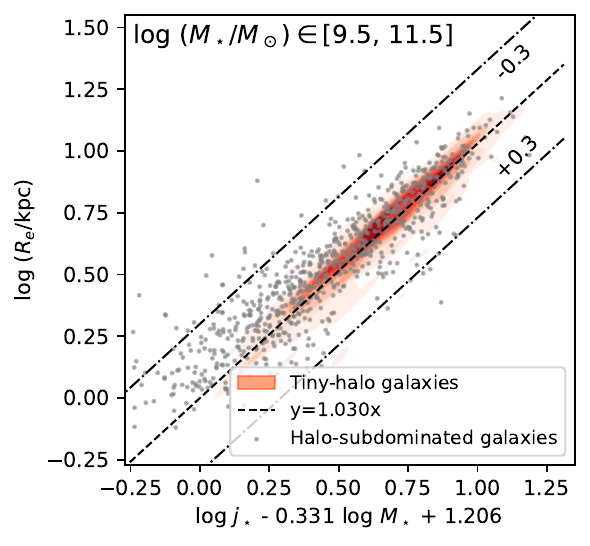}
\includegraphics[width=0.48\textwidth]{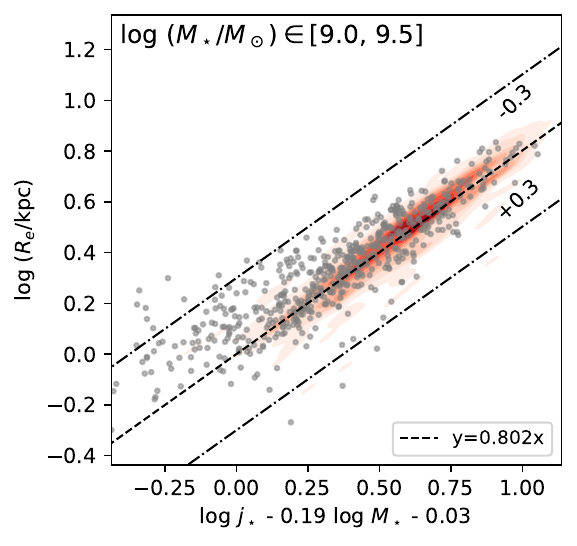}
\caption{The fiducial $j_\star$-$M_\star$-$R_e$ relation of \BD\ galaxies (red KDE contours) from the TNG50 simulation. 
This tight relation is obtained by a surface fitting of the 3D space of $j_\star$, $M_\star$, 
and $R_e$ using two mass ranges ${\rm log}\ (M_\star/M_\odot) \in [9.5,11.5]$ and 
$\in [9.0, 9.5]$. Halo-subdominated galaxies (gray dots) are also shown for comparison. 
\label{fig:fp2}}
\end{center}
\end{figure*}

Moreover, the significance of rotation in disks quantified by $\epsilon f_{\rm disk}$ here is hard to be accurately approximated in observations. 
As shown in \reffig{fig:fdisk}, the mass fraction of morphologically decomposed disk structures $f_{\rm disk, morph}$ are generally slightly 
larger by about $0-0.2$ (median at $\sim 0.1$) in \BD\ galaxies (blue) than those measured using the kinematical method $f_{\rm disk, kinem}$ from 
\citet{Du2019, Du2020}. The difference is larger in halo-subdominated galaxies (red) reaching about $0.05-0.3$ (red, median at $\sim 0.2$).
If we take the potential overestimation of $\epsilon f_{\rm disk}$ from 0.5 to 0.72 in observations into account, it will lead to an offset 
toward the right side of the fiducial \jM-$h_R$ relation about 0.15 dex. The inconsistency between \citet{Pina2021} and TNG50 is understandable. 

In conclusion, we do not find any clear evidence of incorrect properties in the size and rotations of galaxies in TNG50. The inconsistency in 
the \jM-$h_R$ relation between observations and the TNG simulations is likely attributed to several factors:
(1) The substantial uncertainty in the measurement of $j_\star$; (2) overestimation of $\epsilon f_{\rm disk}$; (3) The limited statistical 
quality of the data sample; (4) The bias towards compact galaxies; (5) The contamination from halo-subdominated galaxies. This outcome may further 
lead to the weak dependence on $j_\star$ of the mass-size relation. 

\begin{figure*}[ht!]
\begin{center}
\includegraphics[width=0.48\textwidth]{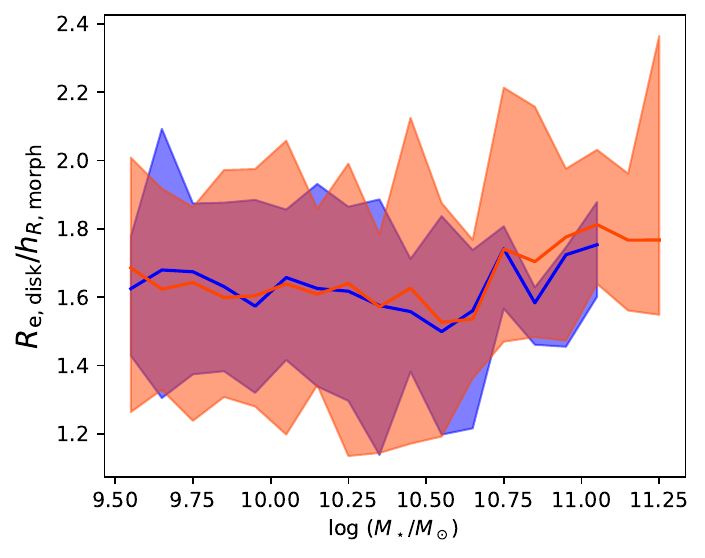}
\includegraphics[width=0.48\textwidth]{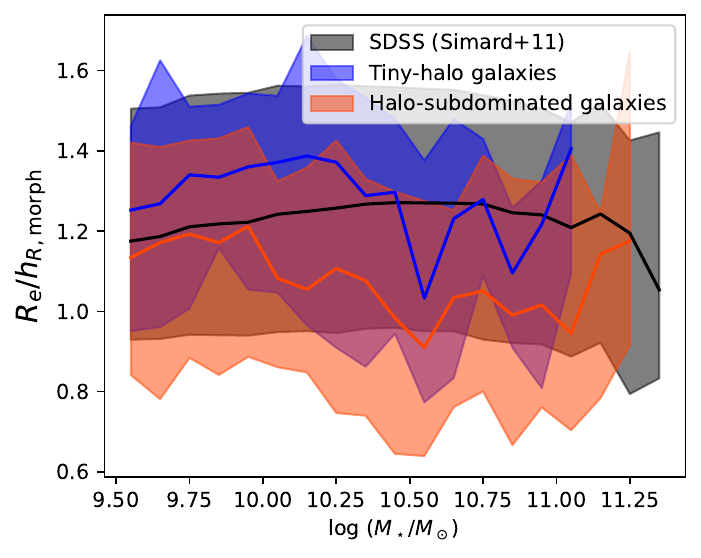}
\caption{The ratio between the effective radius ($R_e$) and the morphologically measured scale 
length ($h_{R, {\rm morph}}$) as a function of stellar mass ($M_\star$). Here, $R_e$ 
represents the half-mass radii of entire galaxies, while $R_{e,{\rm disk}}$ specifically denotes the half-mass 
radii of the disk structures derived through the kinematic method. Tiny-halo and halo-subdominated galaxies are shown 
in blue and red, respectively. The envelopes of shaded regions represent the 16th and 84th percentiles 
and the solid profile is the median value. The black profile shows the measurement from SDSS $r$-band 
adopted from \citet{Simard2011} for comparison.
\label{fig:RehR}}
\end{center}
\end{figure*}

\begin{figure}[htp]
\begin{center}
\includegraphics[width=0.48\textwidth]{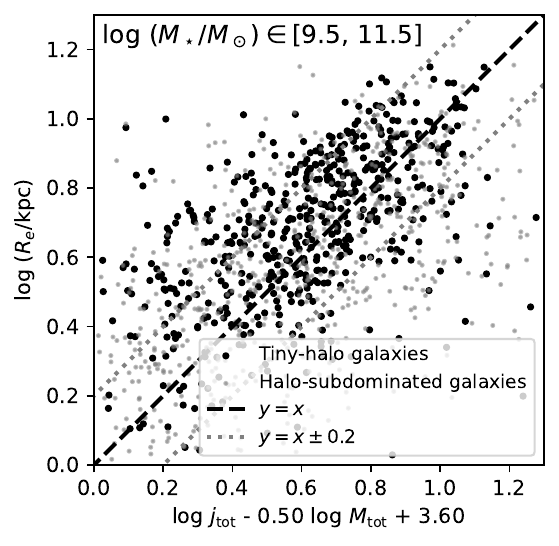}
\caption{The relation between galaxy size and dark matter halo virial radius in the TNG50 simulation. Central \BD\ galaxies (black dots) and central halo-subdominated galaxies (gray dots) are shown here. The $x$-axis represents $\lambda r_{\rm vir}$. This correlation is given by \refeq{eq:jMdm_Re} instead of running a 3D surface fitting.
\label{fig:jMRe_dm}}
\end{center}
\end{figure}

\section{The \jMRe\ relation: the origin of the mass-size relation and its scatter}\label{sec:jMsize}

\subsection{The \jMRe\ relation}
A similar fiducial \jMRe\ relation can be derived, assuming that the half-mass radius of galaxy $R_e$ is proportional to $h_R$. 
Figure \ref{fig:fp2} shows the surface fitting result of the 3D space of $j_\star$, $M_\star$, 
and $R_e$ using two mass ranges. The fitting result of \BD\ galaxies with $M_\star\in [10^{9.5}, 10^{11.5}]M_\odot$ 
(red KDE map in the left panel of \reffig{fig:fp2}) gives 
\begin{equation}\label{eq:MjRe}
\begin{split}
    \mathrm{log}\ &R_e = (1.019\pm 0.015)\ [\mathrm{log}\ j_\star \\ 
                    & -(0.331\pm 0.012)\ \mathrm{log}\ M_\star + (1.206\pm 0.093)].
\end{split}
\end{equation} 
It is clear that $R_e$ follows a very similar correlation to the fiducial \jM-$h_R$ relation in \reffig{fig:fp3}. 
This equation can also be written as 
\begin{equation}
\begin{split}
    \mathrm{log}\ R_e \simeq \mathrm{log}\ j_\star -0.29\ \mathrm{log}\ M_\star\
                     + (1.2-0.04\ \mathrm{log}\ M_\star)
\end{split}
\end{equation}
that can be directly compared with the fiducial \jM-$h_R$ relation. The constant term $1.2-0.04 \log (M_\star/M_\odot)$ is 
approximately 0.8 for the sample of galaxies whose $\log (M_\star/M_\odot) \simeq 10$. Remarkably, the \BD\ galaxies exhibit 
a tight correlation between \jM\ and galaxy size, which closely resembles the \jM-$h_R$ relation. We have verified that 
the half-mass radius of kinematically-derived disk structures, denoted as $R_{e, {\rm disk}}$, is 1.4-1.8 (median at $\sim 1.6$) 
times larger than $h_{R, {\rm morph}}$ (left panel of \reffig{fig:RehR}). This result 
aligns well with the expected behavior for disk structures with exponential profiles. Furthermore, the ratio between 
$R_e$ and $h_{\rm R, morph}$ $R_e/h_{\rm R, morph}$ varies from 0.9 to 1.5 (median at $\sim 1.2$) in both TNG50 and as 
the scale length $h_R$, shown in \reffig{fig:RehR}. 

\refeq{eq:MjRe} can also be written as $R_e \propto \lambda M_\star^{0.212}$ where the spin parameter 
of galaxies $\lambda \propto \frac{j_\star}{M_\star^{0.543}}$ is nearly constant according to the second panel of 
\reffig{fig:jM}. It thus gives the overall mass-size relation $R_e\propto M_\star^{0.225}$ of disk galaxies with 
$M_\star\in [10^{9.5}, 10^{11.5}]M_\odot$ shown in the first panel of \reffig{fig:jM}. The large scatter originates 
from the scatter of $\lambda$. Therefore, the fiducial \jMRe\ relation explains well both the mass-size relation of 
disk galaxies and its large scatter. This result suggests that the disk structure of galaxies while 
displaying a broad range of sizes, does not deviate significantly from the exponential hypothesis. Stellar migration 
\citep[e.g.,][]{Debattista2006, Roskar2012, Berrier&Sellwood2015} does not dramatically alter the fiducial \jM-size relation. 
It is also worth emphasizing that halo-subdominated galaxies (gray dots in \reffig{fig:fp2}) adhere to a scaling relation that is nearly 
identical to \BD\ galaxies, despite the large scatter. We can conclude that nature shapes the overall \jM\ and mass-size relation.
And the effect of external factors plays a minor role. 

\subsection{The relation between galaxy size and dark matter halo virial radius}

We adopt ${\rm log}\ j_\star \simeq {\rm log}\ j_{\rm tot}$ and ${\rm log}\ M_\star \simeq ({\rm log}\ M_{\rm tot}-4.8)/0.67$ 
from \citet{Du2022}, then \refeq{eq:MjRe} can be written as 
\begin{equation}\label{eq:jMdm_Re}
\begin{split}
    {\rm log}\ R_e & \simeq {\rm log}\ j_{\rm tot} - 0.50 \ {\rm log}\ M_{\rm tot} + 3.60
\end{split}
\end{equation}
Such a correlation does exist for galaxies with stellar mass $10^{9.5}-10^{11.5}\ M_\odot$ but has a quite large 
scatter, as shown in \reffig{fig:jMRe_dm}. Note that we here do not run a surface fitting of $j_{\rm tot}$, 
$M_{\rm tot}$, and $R_e$ due to the bad statistic and large scatter.
When considering $r_{\rm vir}/{\rm kpc} \simeq 0.02 (M_{\rm tot}/M_\odot)^{1/3}$ and the constant spin parameter defined as 
$\lambda_{\rm tot} = (10^{3.60}/0.02)\ j_{\rm tot}/M_{\rm tot}^{0.81}=0.08$ adopting $j_{\rm tot}/M_{\rm tot}^{0.81}\sim 10^{-6.37}$ 
from \citet[][Equation 8]{Du2022}, \refeq{eq:jMdm_Re} is written as
\begin{equation}
\begin{split}
    R_e & \sim \lambda_{\rm tot} r_{\rm vir} \sim 0.08 r_{\rm vir}
\end{split}
\end{equation}
%6370 = 4**(1/3) * 10**3.585 = 10**3.6 * 1.6
As a result, TNG50 predicts that the ratio of galaxy size to halo virial radius is $R_e/r_{\rm vir}\sim 0.08$. 
This finding is in agreement with results derived from pure $N$-body simulations, where $R_e \propto \lambda R_{\rm vir}$ 
\citep{Mo1998, Dalcanton1997, Somerville2008}, while also considering the adjustment for the offset from 
$j\propto M^{2/3}$ and the correction of the stellar-to-halo mass relation for disk galaxies. It's important to 
emphasize that defining $\lambda \propto j/M^{2/3}$ in a conventional manner introduces an additional dependency 
on the mass. The size-size relation is very sensitive to the scaling factor of ${\rm log}\ (M_{\rm tot}/M_\odot)$. A deviation of 
$0.02{\rm log}\ (M_{\rm tot}/M_\odot)$ can lead to an uncertainty of $R_e$ by factor 2. 
Additionally, we have validated that the correlation does not hold for less massive galaxies, which 
is consistent with the findings of \citet{Karmakar2023}. The halo-subdominated galaxies (represented by gray dots) 
also exhibit a relatively large scatter. Hence, such a size-size correlation is not always apparent, explaining 
the somewhat conflicting conclusions shown in \citet{Yang2023} and \citet{Karmakar2023}. Thus, we suggest the galaxy 
size-virial radius relation should not be viewed as conclusive evidence of whether the characteristics of galaxies are 
dependent on their parent dark matter halos.

\section{Summary}\label{sec:summary}

In this study, we elucidate the influence of internal and external processes on the scaling relations of the specific angular momentum $j_\star$, 
mass $M_\star$, and size of galaxies from TNG50. We employ a fully automatic kinematical method to decompose the kinematic structures of 
\TNG\ galaxies. Galaxies with more massive kinematic stellar halos generally have experienced a stronger influence from external factors, e.g., mergers 
or close tidal interactions with neighbor galaxies.

Our analysis verifies the crucial role played by the inherent scatter in $j_\star$ arising from internal (natural) processes including but not limited to 
the properties of protogalaxies, secular processes, and host dark matter halos of galaxies. We select galaxies that have tiny 
kinematic stellar halos of mass ratio $f_{\rm halo}\leq 0.2$ to isolate the effect of internal physical processes. Such galaxies populate widely 
over the observed \jM\ relation and the mass-size relation in the local Universe. We confirmed that the disk structures of \BD\ galaxies in \TNG\ 
adhere to the exponential hypothesis. The substantial scatter in the \jM\ relation then provides a robust explanation for 
the fiducial \jM-$h_R$ relation. It further leads to the mass-size relation and the large scatter of galaxy size. 
Additionally, our findings indicate that the impact of stellar migrations, as suggested by previous studies, has a minor effect on the 
overall properties of galaxies. The companion piece to this paper will explore the evolutionary process of galaxies of different sizes 
(Ma, Du, et al., in preparation).

Halo-subdominated galaxies with $0.2<f_{\rm halo}\leq 0.5$ are moderately influenced by external processes. Such galaxies closely align with a scaling 
relation similar to that of tiny-halo galaxies, but have a large scatter and systematically offset toward the low $j_\star$ side. 
This result underscores the dominant role of internal factors in shaping the overall \jM\ and mass-size relation, 
with external effects playing a minor role. Additionally, we examine the correlation between galaxy size and the virial radius of the 
dark matter halo after taking into the adjustment for the offset from $j\propto M^{2/3}$ and the correction of the stellar-to-halo 
mass relation for disk galaxies. Such a correlation is likely to be unclear to make a robust conclusion.

\begin{acknowledgements}
The authors acknowledge the support by the Natural Science Foundation of Xiamen, China (No. 3502Z202372006), the
Fundamental Research Funds for the Central Universities (No. 20720230015), the Science Fund for Creative Research Groups of the National Science Foundation of China (NSFC) (No. 12221003), and the China Manned Space Program. 
We also acknowledge constructive comments and suggestions from H. Mo and D.Y. Zhao.
L.C.H. acknowledges the support by the NSFC grant (11721303, 11991052, 12011540375, 12233001), the National 
Key R\&D Program of China (2022YFF0503401), and the China Manned Space Project (CMS-CSST-2021-A04, CMS-CSST-2021-A06). S.L. acknowledges 
the support by the NSFC grant (No. 11988101) and the K. C. Wong Education Foundation. Y.J.P. acknowledges the support by the NSFC Grant No. 12125301, 12192220, 12192222, and the science research grants from the China Manned Space Project with NO. CMS-CSST-2021-A07.
The TNG50 simulation used in this work, one of the flagship runs of the IllustrisTNG project, has been run on the HazelHen Cray XC40-system at the High Performance Computing Center Stuttgart as part of project GCS-ILLU of the Gauss centers for Supercomputing (GCS). This work is also strongly supported by the Computing Center in Xi'an. 
\end{acknowledgements}

\bibliography{Reference_lib}
\bibliographystyle{aasjournal}

\nolinenumbers
\end{document}